\documentclass[aps,prl,amsmath,amssymb,twocolumn,showpacs,showkeys,superscriptaddress]{revtex4-1}

\usepackage[english]{babel}
\usepackage{graphicx}
\usepackage{dcolumn}
\usepackage{bm}
\usepackage{color}
\usepackage{calrsfs}
\DeclareMathAlphabet{\pazocal}{OMS}{zplm}{m}{n}

\begin{document}
\selectlanguage{english}

\title{Polarization anisotropy of the emission from type-II quantum dots}

\author{P. Klenovsk\'y}
\email[]{klenovsky@physics.muni.cz}
\affiliation{Central European Institute of Technology, Masaryk University, Kamenice 753/5, 62500~Brno, Czech~Republic}
\affiliation{Department of Condensed Matter Physics, Faculty of Science, Masaryk University, Kotl\'a\v{r}sk\'a~2, 61137~Brno, Czech~Republic}

\author{D. Hemzal}
\affiliation{Central European Institute of Technology, Masaryk University, Kamenice 753/5, 62500~Brno, Czech~Republic}
\affiliation{Department of Condensed Matter Physics, Faculty of Science, Masaryk University, Kotl\'a\v{r}sk\'a~2, 61137~Brno, Czech~Republic}

\author{P. Steindl}
\affiliation{Central European Institute of Technology, Masaryk University, Kamenice 753/5, 62500~Brno, Czech~Republic}
\affiliation{Department of Condensed Matter Physics, Faculty of Science, Masaryk University, Kotl\'a\v{r}sk\'a~2, 61137~Brno, Czech~Republic}

\author{M. Z\'ikov\'a}
\affiliation{Institute of Physics CAS, Cukrovarnick\'a 10, Praha 6, 162 00, Czech~Republic}

\author{V. K\v{r}\'apek}
\affiliation{Central European Institute of Technology, Brno University of Technology, Technick\'a 10, 61600 Brno, Czech~Republic}

\author{J. Huml\'i\v{c}ek}
\affiliation{Central European Institute of Technology, Masaryk University, Kamenice 753/5, 62500~Brno, Czech~Republic}
\affiliation{Department of Condensed Matter Physics, Faculty of Science, Masaryk University, Kotl\'a\v{r}sk\'a~2, 61137~Brno, Czech~Republic}

\date{\today}

\begin{abstract}
We study the polarization response of the emission from type-II GaAsSb capped InAs quantum dots. The theoretical prediction based on the calculations of the overlap integrals of the single-particle states obtained in the $\vec{k}\cdot\vec{p}$ framework is given. This is verified experimentally by polarization resolved photoluminescence measurements on samples with the type-II confinement. We show that the polarization anisotropy might be utilized to find the vertical position of the hole wavefunction and its orientation with respect to crystallographic axes of the sample. A proposition for usage in the information technology as a room temperature photonic gate operating at the communication wavelengths as well as a simple model to estimate the energy of fine-structure splitting for type-II GaAsSb capped InAs QDs are given.
\end{abstract}


\maketitle
%
%
%
%
The ground state wavefunction of holes in type-II InAs quantum dots (QDs) with $\mathrm{GaAs_{1-y}Sb_y}$ capping layer (CL) resides outside of the dot volume and in general has the form of two mutually perpendicular pairs of segments~\cite{klenovsky10,KleJOPCS}. Remarkably, the pair oriented along [110] crystallographic direction is positioned close to the QD base, while the other, oriented along [1-10], is located above the dot~\cite{klenovsky10,KleJOPCS,Hsu,Ulloa2,FSSGaAsSbVK}. Both the vertical position and the orientation of the hole wavefunction have been recently discussed in the literature~\cite{Jin,LiuSteer,Ulloa2,UlloaHomogSRL,Hsu,FSSGaAsSbVK}. It has been shown elsewhere~\cite{klenovsky10,FSSGaAsSbVK} that the potential minimum for holes in this system results from a delicate interplay between quantum size effect and the piezoelectric potential and is, thus, sensitive to the thickness $d$ of CL~\cite{FSSGaAsSbVK,Hsu}. In this work, we utilize the $d$-dependence of
the orientation of the hole wavefunction to show that the anisotropy of the polarization of the photoluminescence (PL) from this system can be used to determine the vertical position of the hole, as a gated source of radiation with defined polarization, or to estimate the fine-structure splitting energy of the exciton in type-II heterostructure.\\
%
%
%
%
\textit{Theory}.--~To assess the properties of PL from type-II QDs, we have first calculated the one-particle wavefunctions of the InAs QDs capped by the $\mathrm{GaAs_{1-y}Sb_y}$ CL as a function of $d$ by the envelope function approach using the nextnano++ simulation suite~\cite{next} with the inclusion of the elastic strain and piezoelectricity.
Secondly, the oscillator strengths $I_{\mathrm{ab}}$ of the transitions between states $a$ and $b$ were calculated by a customarily built code based on the approach outlined in Refs.~\onlinecite{t_stier,Stier1999},
\begin{equation}
\label{eq_TMEcalc}
I_{\mathrm{ab}}\approx\frac{2m_0}{\hbar^2E_{\mathrm{ab}}}\left|\left<\psi_a\right|\vec{e}\cdot \hat{P}(\vec{r})\left|\psi_b\right>\right|^2,
\end{equation}
where $\psi_a$ and $\psi_b$ are the wavefunctions of the states $a$ and $b$, respectively, and $\psi=\sum_{i=1}^8 \phi_i u_i$ where $\phi_i$ and $u_i$ are the envelope and Bloch functions for the band $i$, respectively. The other quantities are as follows: $\vec{e}$ denotes the polarization vector, $E_{\mathrm{ab}}$ the difference between the eigenenergies of the states $a$ and $b$, $\hat{P}(\vec{r})$ the momentum operator, $m_0$ is the free electron mass, and $\hbar$ the reduced Planck constant.
Furthermore, because of faster spatial variations of $u$ compared to $\phi$ we make the following approximation $\hat{P}(\vec{r})\psi = (\hat{P}(\vec{r})u) \phi + u (\hat{P}(\vec{r}) \phi ) \approx (\hat{P}(\vec{r})u) \phi$. For more details see Refs.~\onlinecite{t_stier,Stier1999,Takagahara2000}.\\
\textit{Results}.--
We focus on the investigation of the polarization anisotropy of interband transitions, i.e~$\{a,b\}=\{c,v\}$, where $c$ ($v$) is the conduction (valence) band state.
Including spin, the exciton transitions are formed by nearly degenerate quadruplets (bright and dark doublets)~\cite{Bayer2002, Bayer1999} which were, however, not resolved in our experiment. Thus, we calculate $I = 1/4\sum_{\{\uparrow,\downarrow\}\otimes\{\uparrow,\downarrow\}}I_{\mathrm{eh}}$ only.\\
In our experiments, both the excitation and the detected PL radiation propagate perpendicularly to the sample surface and PL is thus polarized parallel to that; the angle between the crystallographic direction [110] and the polarization vector is denoted $\alpha$. We plot our theoretical and experimental results in terms of the degree of polarization $C(\alpha)=(I(\alpha)-I_{\mathrm{min}})/(I_{\mathrm{max}}+I_{\mathrm{min}})$ where $I_{\mathrm{min}}$ and $I_{\mathrm{max}}$ are the extremal values of $I(\alpha)$. Note that for the angle $\alpha_{\mathrm{max}}$ such that $I(\alpha_{\mathrm{max}})=I_{\mathrm{max}}$ the previous relation gives $C(\alpha_{\mathrm{max}})\equiv C_{\mathrm{max}}$, the maximum obtained degree of polarization.\\ 
The composition of $\mathrm{GaAs_{1-y}Sb_y}$ CL was kept spatially constant with Sb content of $y=0.24$ in all calculations and only $d$ was varied. Notice that we have chosen the particular value of $y$ to ensure that, except for the thinnest CLs, the transition would be purely of type-II~\cite{klenovsky10,Hsu}.\\
We have calculated the polarization properties
for several dot geometries and In composition profiles, see Tab.~\ref{tab_simulStruct}.
\begin{table}[!ht]
\begin{tabular}{|c|c|c|c|c|}
\hline
shape& $l$ (nm)& $h$ (nm)& aspect& In composition\\
\hline
lens~\cite{AliceSemiell,KleJOPCS,FSSGaAsSbVK}& 12& 4& 0.33& const. In=1\\
pyramid~\cite{Ulloa}& 24& 8& 0.33& const. In=0.6\\
pyramid~\cite{klenovsky10,Offermans}& 24& 8& 0.33& trumpet-like\\
trunc. cone& 24& 6.5& 0.27& const. In=0.8\\
trunc. pyr.~\cite{Hsu}& 24& 6.5& 0.27& const. In=0.8\\
\hline
\end{tabular}
\caption{Properties of the simulated QDs along with the corresponding references.
The parameters $l$ and $h$ denote the base length and height of QD, respectively. The aspect ratio is $h/l$. For the details of the trumpet-like In composition see Ref.~\onlinecite{suppl_prl_generic}.
\label{tab_simulStruct}}
\end{table}
\begin{figure}[!ht]
\begin{center}
\includegraphics[keepaspectratio=true,width=0.4\textwidth]{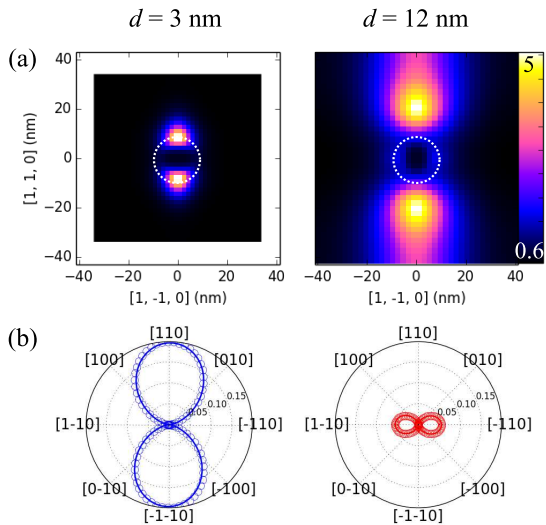}
\end{center}
\caption{(color online) (a) Horizontal cross-section of the normalized hole probability density for the lens-shaped QD with the base diameter and height of 12 and 4~nm, respectively, 
and for CL thicknesses of $d=3$~nm, and $d=12$~nm. The color scale in (a) is in units of $10^{-5}$m$^{-3}$. The cuts cross the position of the maximum of the probability density at 1.8~nm (6.0~nm) above the dot base for CL thickness of 3~nm (12~nm). Panels (b) show the calculated $C(\alpha)$ for the ground state electron-hole transition; full curve is guide to the eye. Dotted white circles in (a) mark the edges of the dot. The electron wavefunction occupies the dot volume and is slightly elongated in the [1-10] direction.
\label{fig1_whatIsCalc}}
\end{figure} 
The choice of the QD shapes was motivated by the results of Refs.~\onlinecite{AliceSemiell,KleJOPCS,FSSGaAsSbVK,Ulloa,klenovsky10,Offermans,Hsu}. Except for the structure with trumpet-like profile~\cite{Offermans,Ulloa,klenovsky10,KleJOPCS}, the values of constant In content in the QDs were chosen so as the ground state emission energy lies within the range of the communication bands, 1.3--1.55$\,\mu$m. Based on the predicted properties of these structures, the technology of QD sample preparation was subsequently optimized to obtain QDs with desired properties, i.e.~the long emission wavelength and type-II band alignment.\\
Typical results of our calculations are shown in Fig.~\ref{fig1_whatIsCalc} for the lens-shaped QD with the base diameter and height of 12 and 4~nm, respectively,
and two CL thicknesses of $d=3$~nm, and $d=12$~nm. Notably,
the emitted light is preferentially polarized 
along [110] ([1-10]) crystallographic direction for thin (thick) CL. 
The dominant contribution to $I(\alpha)$ in QDs
comes from the areas with the largest overlap between electrons and holes. The electron is firmly bound in the body of type-II QDs for all Sb contents and CL thicknesses~\cite{klenovsky10,Hsu}. Thus, also due to the larger effective mass of holes the overlap and the resulting polarization anisotropy of $I(\alpha)$ is dictated by those parts of the hole wavefunction which are located closest to the dot, see panel (a) of Fig.~\ref{fig1_whatIsCalc}.\\
\begin{figure}[!ht]
\begin{center}
\includegraphics[keepaspectratio=true,width=0.45\textwidth]{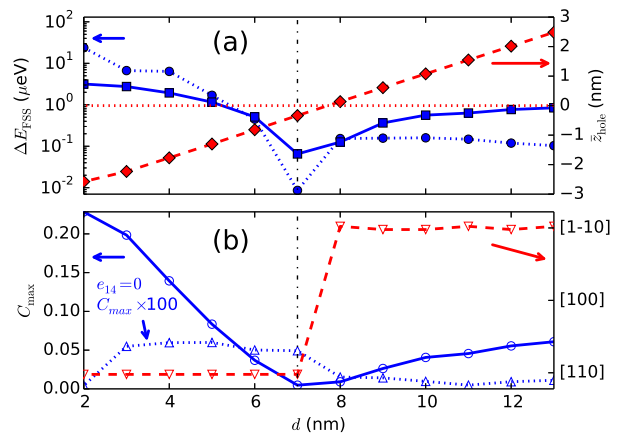}
\end{center}
\caption{(color online) (a) Calculated
$d$-dependencies of $\Delta E_\mathrm{FSS}$ estimated from $C_{\mathrm{max}}$ (full squares), that calculated by the model of Ref.~\onlinecite{FSSGaAsSbVK} (full circles), and the relative vertical position of the hole $\bar{z}_{\mathrm{hole}}$ (full diamonds); (b) $d$-dependencies of $C_{\mathrm{max}}$ (open circles), that with piezo switched off and multiplied by a factor of 100 (upward open triangles), and $\alpha_{\mathrm{max}}$ (downward open triangles). The latter is shown in the crystallographic axes of the QD material. Calculations were done for lens-shaped QD, see Tab.~\ref{tab_simulStruct}. The vertical dot-dash line indicates $d_c$.
The remaining curves are guides to the eye.
\label{fig_calcResults}}
\end{figure}\\
We show in Fig.~\ref{fig_calcResults}~(a) that with $d$ increasing, the vertical position of the hole wave function is first shifted upwards and the average of the vertical hole coordinate measured from the top of the QD, denoted $\bar{z}_{hole}$, increases.
At the same time $C_{\mathrm{max}}$ is reduced [Fig.~\ref{fig_calcResults}(b)] until, for certain critical CL thickness $d_c$, the emission becomes isotropic ($C_{\mathrm{max}}=0$) and the average hole vertical position coincides with the top of the QD ($\bar{z}_{hole}=0$).
During this period also $\alpha_{\mathrm{max}}=0^\circ$ holds and the hole segments are oriented along [110] direction. However, further increase of $d$ results in increase of $\bar{z}_{\mathrm{hole}}$ and $C_{\mathrm{max}}$.
Also a flip of the orientation $\alpha_{\mathrm{max}}$ to [1-10] direction for $d_c$ is observed and this is attained for even larger values of $d$. While $\alpha_{\mathrm{max}}$ is identical with the orientation of the dominant pair of segments of the hole wavefunction, i.e.~those with the largest overlap with the electrons, $C_{\mathrm{max}}$ corresponds to the ratio of overlaps of both pairs with them, see also Fig.~\ref{fig1_whatIsCalc}~(a). The correspondence of $C_{\mathrm{max}}$ is, however, not exact because it is influenced also by a slight elongation of the electron wavefunction, occurring along [1-10].\\
The overlap of electrons and holes is rather small in type-II QDs~\cite{klenovsky10,KleJOPCS,SiGeKlenovsky}. However, particularly for GaAsSb capped InAs QDs we can take advantage of that to estimate the energy of the fine-structure splitting ($\Delta E_\mathrm{FSS}$), i.e.~the energy separation of the bright excitonic doublet, from $C_{\mathrm{max}}$. Motivated by the results of Ref.~\onlinecite{FSSGaAsSbVK} we estimate $\Delta E_\mathrm{FSS}$ as
\begin{equation}
\label{eq_approxFss}
\Delta E_\mathrm{FSS}\sim\frac{e^2}{2\pi\varepsilon\bar{l}_{QD}^3}C_{\mathrm{max}},
\end{equation}
where $e$ is the elementary charge, $\varepsilon$ the permittivity, and $\bar{l}_{QD}$ the mean diameter of the QD base. The length $\bar{l}_{QD}$ replaces $|\vec{r}_{e}-\vec{r}_{h}|$, where $\vec{r}_{e}$ ($\vec{r}_{h}$) is the position of the electron (hole). In constructing $\bar{l}_{QD}$ we have noted a typical situation for type-II GaAsSb capped InAs QDs where the maximum of probability densities for electron and hole is positioned at similar distances from the edge of the dot. To make the estimation even easier we considered the QD to have a shape of a hemisphere with electron located in the center of its circular base. Hence $\bar{l}_{QD}$ is in turn equal to the diameter of the hemisphere's base.
Equation~\ref{eq_approxFss} provides a simple, rough estimate of $\Delta E_\mathrm{FSS}$ from the measured values of $C_{\mathrm{max}}$ and the structural parameters of QDs. The latter can be obtained by standard structural characterization tools. The comparison of Eq.~\ref{eq_approxFss} to more elaborate calculations of $\Delta E_\mathrm{FSS}$, see Ref.~\onlinecite{FSSGaAsSbVK} for details, is shown in Fig.~\ref{fig_calcResults}~(a). The error is as large as an order of magnitude. Evidently, Eq.~\ref{eq_approxFss} gives most precise estimates of $\Delta E_\mathrm{FSS}$ for small values of $C_{\mathrm{max}}$. The value of $\alpha_{\mathrm{max}}$ is identical with that for the lower energy exciton of the corresponding exciton doublet.\\
We have calculated $\alpha_{\mathrm{max}}$ and $C_{\mathrm{max}}$ also for other QD shapes and In composition from Tab.~\ref{tab_simulStruct} and confirmed the existence of $d_c$ associated with the vertical position of the (ground state) hole wavefunction reaching the QD apex also for pyramidal QDs. Only in truncated shapes the hole segments are oriented along [110] for all $d$ values in agreement with the results of Ref.~\onlinecite{Hsu}. We note that the error of the estimate of $\Delta E_\mathrm{FSS}$ was the same for all studied QDs listed in Tab.~\ref{tab_simulStruct}.\\
It is the correspondence between the orientation of its segments and the vertical position of the hole wavefunction in respect to the QD volume that enables us to determine its position for untruncated QD shapes in real experiments simply by measuring the direction of the in-plane PL polarization anisotropy.\\
The behavior just described is closely connected to the piezoelectricity induced by the shear stress as can be seen from the dependence of $C_{\mathrm{max}}$ on $d$ being more than hundred times weaker when the piezoelectric term $e_{14}$ is set to zero in the calculations, see the dash-dotted curve in Fig.~\ref{fig_calcResults}~(b). We note that the effect of the shear elements of the Bir-Pikus strain Hamiltonian~\cite{BirPik} via the deformation potentials is negligible as well.\\
The quantities $C_{\mathrm{max}}$ and $\alpha_{\mathrm{max}}$ have an intimate relation to the mixing of heavy (HH) and light (LH) hole states as that is responsible for the shape and orientation of the envelope wavefunction of the hole~\cite{kleDresden,Musial2012}. For our dots the phase of HH-LH mixing determines $\alpha_{\mathrm{max}}$, and its amplitude $C_{\mathrm{max}}$. Both parameters can be determined using the results of Refs.~\onlinecite{tonin12,leger07}.\\
%
%
%
%
To test our theoretical predictions, three samples of InAs QDs with GaAsSb CL, denoted A, B and C, were fabricated. In view of the possible future applications, the low pressure metal-organic vapor phase epitaxy (MOVPE) growth was chosen~\cite{Hospodkova2010surfSci}. We have prepared the samples so that the amount of Sb in CL would correspond to type-II band alignment, which was confirmed by the blue shift of PL with increased pumping, see Ref.~\onlinecite{suppl_prl_generic}. The InAs QD layer was grown under different V/III ratios for each sample. More information about sample preparation can be found in the supplement, Ref.~\onlinecite{suppl_prl_generic}. Different conditions of sample preparation along with the nonuniform growth on the non-rotating susceptor provided us with required variety of QDs with different sizes, shapes, and CLs with variable Sb contents and thicknesses.\\
The measurements of the polarization anisotropy of PL were performed using the NT-MDT Ntegra-Spectra spectrometer. The samples were positioned in the cryostat and cooled to liquid nitrogen temperature (LN2) (for room temperature PL results see Ref.~\onlinecite{suppl_prl_generic}) and were pumped by the solid-state laser with the wavelength of~785~nm and maximum power on the sample surface of 30~mW which was varied by a tunable neutral density (ND) filter.
In every experiment we have collected PL light from large number of QDs ($\sim$2000)
in order to damp the deviations from the mean properties of the ensemble of dots. The polarization of PL was analyzed by a rotating half-wave plate followed by a fixed linear polarizer. Finally, the PL signal was dispersed by a 150 grooves/mm ruled grating and detected by the InGaAs line-CCD camera, cooled to minus 90~$^\circ$C.\\
\begin{figure}[!ht]
\begin{center}
\includegraphics[keepaspectratio=true,width=0.45\textwidth]{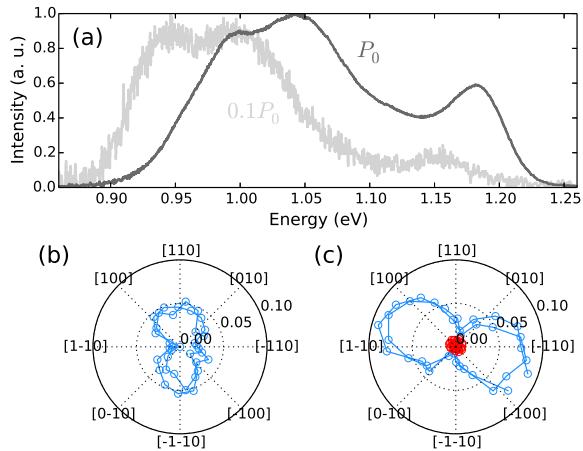}
\end{center}
\caption{(color online) (a): PL spectra measured at LN2 temperature of sample C for pumping powers of $P_0=$14 kWcm$^{-2}$ (dark gray curve) and 1.4 kWcm$^{-2}$ (light gray curve). Notice the blueshift of 50 meV between the spectra.
(b) and (c): polarization dependence of 
the intensity of the lowest-energy PL band for samples A and C, respectively (blue circles). The polarization angles are represented by the corresponding crystallographic axes. Note that the lowest transitions are polarized perpendicularly from one sample to the other. The red circles represent the residual polarization anisotropy of the setup which was 0.005. In (b) and (c) the lines are guides to the eye.
\label{fig_measurementOutline}}
\end{figure}
The experimental procedure was as follows. First, the type of confinement was determined by measuring the pumping dependency of the PL spectra. Either a blue-shift of the whole spectrum was observed, indicating type-II confinement~\cite{Jin,SiGeKlenovsky}, or the spectrum only increased in magnitude without any visible spectral shift, and we considered it as type I. For a typical example of the results obtained on sample C see Fig.~\ref{fig_measurementOutline}~(a). Consequently, the PL spectra for 37 angular positions of $\lambda/2$-plate at fixed laser pumping power were acquired and fitted by a sum of Gauss-Lorentz (GL) profiles~\cite{Hum}. As our calculations described above are most accurate for the ground-state transition, we consider here the intensity of the GL band with the lowest energy only. Examples are shown in Fig.~\ref{fig_measurementOutline}~(b) and (c) for type-II samples A and C. Note that the dominant polarization direction is clearly oriented along [110] in (b), and close to [1-10] in (c). We stress that due to expected small values of $C_{\mathrm{max}}$ we paid a particular attention to (i) the reduction of the residual polarization of the whole setup, see red circles in Fig.~\ref{fig_measurementOutline}~(c), and (ii) proper fitting of the spectra.
\begin{figure}[!ht]
\begin{center}
\includegraphics[keepaspectratio=true,width=0.45\textwidth]{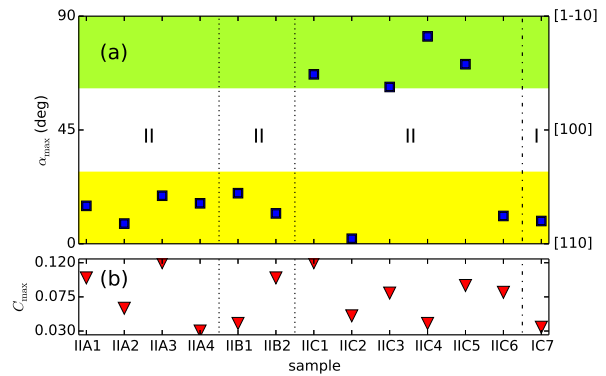}
\end{center}
\caption{(color online) (a): Measured $\alpha_{\mathrm{max}}$ reduced to the first quadrant (blue squares), and (b) $C_{\mathrm{max}}$ (red triangles) for three samples (A--C) and different positions on them, labeled by $xyz$, where $x$ refers to the type of confinement, $y$ to the sample and $z$ to the position on it. The dashed vertical lines mark different samples and the dash-dotted one different type of confinement which is also denoted by I or II in panel (a). The green/yellow stripes in (a) are the typical errors of $\alpha_{\mathrm{max}}$ of 30$\,$deg. The right vertical axis in (a) represents the crystallographic axes of the sample.
\label{fig_measuredResults}}
\end{figure}\\
The resulting $\alpha_{\mathrm{max}}$ for all samples and different positions on them are summarized in Fig.~\ref{fig_measuredResults}~(a). For type II we have observed $\alpha_{\mathrm{max}}$ corresponding to both [110] and [1-10] crystallographic directions, the former being more frequent. This in turn means that holes were located preferentially close to the base of the QD. Furthermore, for type I $\alpha_{\mathrm{max}}$ corresponded to [110] direction, in agreement with the results on InAs/GaAs QDs~\cite{HumPhysE}. We note, however, that the spread of the obtained values of $\alpha_{\mathrm{max}}$ might be as large as 30 degrees. We assume that this is an effect of the QD and CL irregularities which were still not averaged out and also due to remaining imperfections of the measurement procedure.\\
The measured values of $C_{\mathrm{max}}$ are given in Fig.~\ref{fig_measuredResults}~(b) and are varying from 0.03 to 0.12. Clearly, type-II GaAsSb capped InAs QDs might possess a significant polarization anisotropy.\\
Finally, the nature of the studied effect opens a number of possibilities to tune $\alpha_{\mathrm{max}}$ and $C_{\mathrm{max}}$ by external fields, e.g.~electric or strain~\cite{klenovsky10,Trotta,kleDresden}.
In the information technology, this could enable the two perpendicular polarizations of the emitted photons to serve as a low-drain photonic realization of ``zeros" and ``ones", with the bonus of operation at communication wavelengths. In addition, this behavior can be reached for large ensembles of QDs, which was demonstrated here by measurements on the MOVPE-prepared samples, and even at room temperature.\\
%
%
%
%
\textit{Conclusions}.--~We have theoretically predicted, and for the first time also experimentally observed two perpendicular PL polarizations of type-II GaAsSb capped InAs QDs and explained it as an almost exclusive effect of the piezoelectricity. The measurement of polarization anisotropy enables the determination of the vertical position of the holes and their orientation in the studied system. Furthermore, a simple relation to estimate the energy of FSS in type II QD structures and the utilization of the polarization anisotropy as a room-temperature gate based on photons with energy in the communication bands and defined polarization state were proposed.\\
\textit{Acknowledgments}.--~The authors thank Alice Hospodkov\'a for her help with the sample preparation and fruitful discussions. The work was supported by the project no. TH01010419 of the Technological agency of the Czech Republic, the European Regional Development Fund, project
No. CZ.1.05/1.1.00/02.0068, and the European Social Fund, grant No. CZ.1.07/2.3.00/30.0005.

\end{document}